\documentclass[letterpaper, 10 pt, conference]{ieeeconf}  % Comment this line out if you need a4paper

\IEEEoverridecommandlockouts                              % This command is only needed if 
\usepackage{multirow}

\usepackage{amsthm}
\usepackage{amsmath, comment} % assumes amsmath package installed
\usepackage{amssymb} 
\usepackage{graphicx}
\usepackage{algorithm} 
\usepackage[noend]{algpseudocode}
\usepackage{xcolor}

\theoremstyle{definition}

\theoremstyle{remark}

\usepackage{caption}
\usepackage{subcaption}
\usepackage{stackengine}

\usepackage{cancel}
\usepackage{booktabs}
\usepackage{makecell}
\usepackage{cite}
\usepackage{placeins}
\usepackage[colorlinks=true,
            linkcolor=blue,
            citecolor=blue,
            urlcolor=blue]{hyperref}

\title{\LARGE\bf An Adjoint-based Neural Regulator for Real-Time Optimal Control with State Constraints}

\author{Isaiah A. Agboola$^{1}$, Yuxin Tong$^{1}$, Uduak Inyang-Udoh$^{1}$ \\
\thanks{$^{1}$Department of Mechanical Engineering, University of Michigan, Ann Arbor, Michigan, USA. \tt\small agboola@umich.edu, \tt\small yuxinton@umich.edu, \tt\small udiinyang@umich.edu} }

\begin{document}
\maketitle
\thispagestyle{empty}
\pagestyle{empty}

\begin{abstract}
This paper introduces a learning-based control framework for real-time constrained optimal control of nonlinear systems with safety guarantees based on the Pontryagin’s Minimum Principle. The approach learns a neural co-state (adjoint) policy that encodes optimality through the system Hamiltonian, rather than directly approximating a control law. Feasibility is enforced separately at runtime through an efficient convex projection that incorporates actuator limits and safety constraints expressed as control barrier functions. We refer to this framework as an adjoint-based neural regulator (ANR) as it yields a controller that satisfies constraints while retaining the optimality structure encoded by the learned adjoint. We demonstrate the effectiveness of the proposed framework on nonlinear constrained control tasks using a unicycle model. The ANR achieves performance at par with nonlinear model predictive control at more than two orders of magnitude lower computational cost, while exhibiting near-invariant performance across unseen scenarios, thus, significantly outperforming reinforcement learning methods in out-of-training-distribution regimes.

\end{abstract}

\makeatletter
\renewcommand{\@makefnmark}{}
\makeatother
\footnotetext{Code available
\href{https://github.com/AI-Sys-Lab/Adjoint-based-Neural-Regulator.git} {here}.}

\section{Introduction} \label{sec-introduction}

Real-time constrained optimal control is a central challenge in robotics and autonomous systems, where controllers must balance performance with constraint satisfaction under limited computational time \cite{hsu2024safety,ames2019cbf}. 
In nonlinear systems, optimal control is most widely implemented using nonlinear model predictive control (NMPC), which solves a finite-horizon optimization problem online \cite{rawlings2020mpc}. %and naturally accommodates constraints 
However, in safety-critical and fast dynamical settings, repeated nonlinear optimization and solver latency can make NMPC prohibitively expensive in practice \cite{diehl2009nmpc,hewson2023real}. 
This computational burden limits its deployability in certain applications where control decisions must be made on millisecond timescales \cite{hewson2023real}.

Learning-based control, particularly reinforcement learning (RL), offers an appealing alternative by amortizing computation into an offline training phase and producing fast online policies \cite{brunke2022safe}. 
Despite impressive empirical successes, RL often faces practical barriers for constrained control problems. 
In particular, RL often requires substantial reward engineering, can become increasingly sample-inefficient over long horizons, may generalize poorly to out-of-distribution conditions, and does not generally provide strict constraint satisfaction without additional safety mechanisms \cite{garcia2015safe, brunke2022safe, 9928337}. 

A common strategy for enforcing safety is to combine RL with control barrier functions (CBFs), projecting a learned nominal action onto the set of safe controls through a quadratic program (QP) \cite{ames2019cbf,cheng2019end}. 
While effective in many settings, this nominal-then-project paradigm introduces structural limitations. 
First, feasibility can become challenging when CBF-based state safety conditions must be enforced simultaneously with nontrivial or coupled input constraints \cite{ames2019cbf}. 
Second, if the nominal policy does not reflect the underlying optimality structure of the control problem, the safety filter may produce conservative or aggressive corrective actions, leading to frequent and large projections \cite{cheng2019end}. These issues arise because the nominal policy does not explicitly encode the optimality structure of the constrained control problem.

To address these limitations, we utilize an adjoint-based neural control framework that, based on prior work in \cite{11312025}, learns adjoint variables rather than control actions directly. These adjoints represent the sensitivity of the value function (control objective) with respect to the state and encode the indirect optimality structure of the control problem \cite{pontryagin1962},\cite{11107540}. By predicting adjoints online, the controller computes the applied input through a real-time constrained Hamiltonian minimization, in which state safety and input feasibility are enforced simultaneously. 
This yields a unified framework for handling hard state constraints through CBF conditions together with input constraints, while preserving the computational advantages of amortized inference. 
We evaluate the proposed approach on a unicycle with safety constraints and compare its performance against NMPC and RL baselines.

This paper is organized as follows. Sec. \ref{sec-background} reviews NMPC and RL-based approaches, along with CBFs, for constrained optimal control, and motivates the adjoint-based learning approach. Sec. \ref{sec-methodology} describes the adjoint-based neural controller while Sec. \ref{sec-example} presents results in contrast to those obtained using NMPC and RL. We conclude in Sec. \ref{sec-conclusion}.

\section{Background and Motivation} \label{sec-background}

\subsection{Constrained Optimal Control}

Consider a continuous-time optimal control problem (OCP) over a fixed horizon $[0,t_f]$:
\begin{subequations} \label{eq:background_ocp_formulation}
\begin{align}
\min_{\mathbf{u}(\cdot)} \quad 
\mathcal{J}
&=
\phi(\mathbf{x}(t_f))
+
\int_{0}^{t_f}
\ell(\mathbf{x}(t),\mathbf{u}(t),t)\,dt,
\label{eq:background_cost}\\
\text{s.t.}\quad
\dot{\mathbf{x}}(t)
&=
f\!\left(\mathbf{x}(t)\right)
+
g\!\left(\mathbf{x}(t)\right)\mathbf{u}(t),
\label{eq:background_dynamics} \quad
\mathbf{x}(0) = \mathbf{x}_0,\\
%\label{eq:background_ic}\\
\mathbf{u}(t) &\in \mathcal{U},
\label{eq:background_input_constraint}\\
h_i(\mathbf{x}(t)) &\ge 0,\quad i=1,\dots,m,\ \forall t\in[0,t_f].
\label{eq:background_state_constraint}
\end{align}
\end{subequations}
Here, $\mathbf{x}(t)\in\mathbb{R}^n$ is the system state and $\mathbf{u}(t)\in\mathbb{R}^q$ is the control input. The admissible input set $\mathcal{U}$ captures actuator limits, and the inequalities $h_i(\mathbf{x}(t))\ge 0$, $i=1,\dots,m$, encode hard state-safety requirements such as obstacle avoidance \cite{huang2023tvcbf}. The functions $\ell(\mathbf{x}(t),\mathbf{u}(t),t)$ and $\phi(\mathbf{x}(t_f))$ denote the running (stage) and terminal costs, respectively. Quadratic costs are typically adopted: \cite{bertsekas1995dynamic}: 
\begin{equation}
\begin{aligned}
    \ell(\mathbf{x}(t),\mathbf{u}(t)) &=  \mathbf{x^\top}(t)Q\mathbf{x}(t) + \mathbf{u^\top}(t)R\mathbf{u}(t), \\ \quad \phi(\mathbf{x}(t_f)) &= \mathbf{x^\top}(t_f)P\mathbf{x}(t_f)
\end{aligned}
\label{eq:stage_terminal_cost}
\end{equation}
where $Q, P\succeq 0$ and $R\succ 0$ are weightings.

\subsection{Nonlinear Model Predictive Control (NMPC)}

A standard approach for handling constrained OCP in Eq.~\eqref{eq:background_ocp_formulation} is NMPC, which, at each sampling instant solves a finite-horizon constrained optimization problem using the current state as the initial condition\cite{rawlings2020mpc} and the discretized system dynamics. At time $t_k$, let $\mathbf{x}_{k+j}$ and $\mathbf{u}_{k+j}$ denote the predicted state and control input at time step $k+j$ over the prediction horizon. The NMPC problem is formulated as:
\begin{subequations}\label{eq:nmpc}
\begin{align}
\min_{\mathbf{u}_{k+j, j =0,\dots,N-1}}\quad 
 \phi(\mathbf{x}_{k+N})
+ \sum_{j=0}^{N-1}\ell(\mathbf{x}_{k+j},\mathbf{u}_{k+j}) \\
\text{s.t.}\quad
\mathbf{x}_{k+j+1}=f_d(\mathbf{x}_{k+j},\mathbf{u}_{k+j}),
~~ j=0,\ldots,N-1, \\
\mathbf{u}_{k+j}\in\mathcal{U},
\quad j=0,\ldots,N-1, \\ 
h_i(\mathbf{x}_{k+j})\ge 0,
\quad i=1,\ldots,m,\;\; j=1,\ldots,N.
\end{align}
\end{subequations}
where $f_d(\cdot)$ denotes the system dynamics in discrete-time. In this formulation,both state and input constraints are enforced directly within the optimization. Its main limitation, however, is that Eq.~\eqref{eq:nmpc} must be solved online at every feedback step as a constrained nonlinear program. For fast dynamical systems or safety critical tasks, this repeated online optimization can become computationally prohibitive.

\subsection{Reinforcement Learning via SAC}

Reinforcement learning (RL) offers an alternative to repeated online optimization by shifting most computation to offline training and enabling fast policy evaluation during deployment. In control applications, soft actor–critic (SAC), is among the most widely used RL algorithms. SAC learns a stochastic control policy that maximizes returns (or the negation of the control objective). Let $\pi_\theta$ denote a nominal stochastic policy parameterized by $\theta$, which generates the control input according to $\mathbf{u_k}\sim \pi_\theta(\cdot\mid \mathbf{x_k}).$ Then, over a distribution of training tasks or environments $\xi \sim \mathcal{D}$, the policy is learned by maximizing the expected entropy-regularized discounted return $r_\xi$ \cite{11018429}:

\begin{equation}
\begin{aligned}
\arg\max_\theta\;
\mathbb{E}_{\substack{\xi \sim \mathcal{D}\\
\tau \sim \pi_\theta}}
\Bigg[
\sum_{k=0}^{T_\xi-1}
\gamma^k
\Big(
r_\xi(\mathbf{x}_k,\mathbf{u}_k)
-
\alpha \log \pi_\theta(\mathbf{u}_k \mid \mathbf{x}_k)
\Big)
\Bigg]
\end{aligned}
\label{eq:bg_rl_dt_obj}
\end{equation}
where $\gamma \in (0,1)$ is the discount factor, $\alpha > 0$ is the entropy regularization weight, and  $\tau$ denotes a rollout generated by executing policy $\pi_\theta$ in training environment $\xi$.

Eq.~\eqref{eq:bg_rl_dt_obj} is evaluated using state-transition samples collected from simulator rollouts and stored in a replay buffer \cite{9928337}. The learned policy provides fast nominal control actions during deployment, but by itself does not guarantee satisfaction of state or input constraints. For this reason, the nominal SAC policy is typically augmented with a safety filter at deployment, as described next.

\subsection{CBF-Based Safety Filtering}

To enforce safety during deployment, the nominal policy output is passed through an online safety filter. For the control-affine dynamics in Eq.~\eqref{eq:background_dynamics} with input constraint set $\mathcal{U}$, safety may be enforced using control barrier functions (CBFs) which may be expressed as \cite{kong2025differential}:
%. When a safety function has relative degree one, the CBF condition can be written as
\begin{equation}
L_f h(\mathbf{x}) + L_g h(\mathbf{x})\,\mathbf{u} + \alpha(h(\mathbf{x})) \ge 0,
\label{eq:cbf_condition}
\end{equation} 
where $\alpha(\cdot)$ is any extended class-$\mathcal{K}$ function, and $L_f h$, $L_g h$ denote the Lie derivatives along $f$ and $g$ respectively. For the obstacle-avoidance, higher-order CBFs (HOCBFs) may be employed, by taking higher Lie derivatives \cite{10768176, xiong2023discrete}. For example, for the state constraint (safety function) in \eqref{eq:background_state_constraint} we may define a conservative state-dependent admissible control set with a HOCBFs of relative degree two as \cite{kong2025differential}
\begin{equation}
\begin{aligned}
\mathcal{S}(\mathbf{x})&
:=
\Big\{ 
\mathbf{u}\in\mathbb{R}^{n_u}
\;\Big|\;
L_f^2 h_i(\mathbf{x})
+
L_gL_f h_i(\mathbf{x})\,\mathbf{u}
 \\
& +
k_{1,i}L_f h_i(\mathbf{x})
+
k_{0,i}h_i(\mathbf{x})
\ge 0,\;
i=1,\dots,m
\Big\}.
\end{aligned}
\label{eq:hocbf_admissible_set}
\end{equation}
where $k_{1,i},k_{0,i}>0$ are constant HOCBF gains. Together with the input constraint set $\mathcal{U}$, this defines the inputs that are both safe and actuator-feasible. Hence, given the nominal SAC policy output \(u^{\mathrm{RL}}(t)\), we apply a HOCBF-QP safety filter that computes the smallest correction required to satisfy the input and safety constraints. The deployed control is
\[
u^*(t)=u^{\mathrm{RL}}(t)+\Delta u^*(t),
\]
where
\[
\Delta u^*(t)
=
\arg\min_{\Delta u}\|\Delta u\|_2^2
\quad
\text{s.t.}\quad
u^{\mathrm{RL}}(t)+\Delta u \in \mathcal U\cap\mathcal S(x(t)).
\]
Equivalently,
\[
u^*(t)
=
\arg\min_{u\in\mathcal U\cap\mathcal S(x(t))}
\|u-u^{\mathrm{RL}}(t)\|_2^2.
\]

While the CBF modifies the RL policy output to enforce safety constraints \cite{janani2025fixed}, the RL-CBF framework still inherits key challenges of RL, including ill-conditioned learning with long-horizon or high system dimensionality, which lead to sample inefficiency and/or poor generalizability \cite{yang2025cbfrl}.

\subsection{PMP Optimality Structure for Adjoint Prediction}

It is known from Pontryagin's Minimum Principle (PMP) that the state, control input pair $({\mathbf{x}}^\star(t), \mathbf{u}^\star(t))$ that solves the OCP in Eq.~\eqref{eq:background_ocp_formulation} must satisfy the so-called two-point boundary value problem (TPBVP):
\begin{align}
\dot{\mathbf{x}}^\star(t)
&=
\nabla_{\boldsymbol{\lambda}} H(\mathbf{x}^\star(t),\mathbf{u}^\star(t),\boldsymbol{\lambda}^\star(t),t),
\label{eq:bg_pmp_state}\\
\dot{\boldsymbol{\lambda}}^\star(t)
&=
-\nabla_{\mathbf{x}} H(\mathbf{x}^\star(t),\mathbf{u}^\star(t),\boldsymbol{\lambda}^\star(t),t),
\label{eq:bg_pmp_costate}\\
\mathbf{u}^\star(t)
&=
\arg\min_{\mathbf{u}\in\mathcal{U}}
H(\mathbf{x}^\star(t),\mathbf{u},\boldsymbol{\lambda}^\star(t),t),
\label{eq:bg_pmp_control}\\
\mathbf{x}^\star(0)&=\mathbf{x}_0,
\qquad
\boldsymbol{\lambda}^\star(t_f)=\nabla \phi(\mathbf{x}^\star(t_f)).
\label{eq:bg_pmp_bc}
\end{align}

where $H$ is the control Hamiltonian defined as

\begin{equation}
\begin{aligned}
H(\mathbf{x}(t),\mathbf{u}(t),\boldsymbol{\lambda}(t),t)
=
\ell(\mathbf{x}(t),\mathbf{u}(t),t) \\
+
\boldsymbol{\lambda}(t)^\top
\Big(
f(\mathbf{x}(t))
+
g(\mathbf{x}(t))\mathbf{u}(t)
\Big),
\label{eq:bg_H}
\end{aligned}
\end{equation}
and $\boldsymbol{\lambda}(t)\in\mathbb{R}^n$ denotes the co-state, or adjoint.

The adjoint captures the local sensitivity of the optimal cost-to-go to the state, while Eq.~\eqref{eq:bg_pmp_control} induces a control action through the Hamiltonian minimization. Although the TPBVP is generally difficult to solve in real time, recent work has shown that its solution can be approximated via neural parameterization of the adjoint \cite{lian2025ncpr}. \textit{This suggests that the RL-CBF paradigm for fast control inference and safety enforcement can be replaced by a framework that parameterizes optimality through the adjoint and enforces feasibility/safety via a constrained Hamiltonian minimization, thereby overcoming the long-horizon control learning limitations in RL.} This framework is developed in the next section.

\section{Adjoint-Based Neural Control}\label{sec-methodology}

\begin{figure*}[t]
    \centering
    
    \begin{subfigure}{0.55\textwidth}
        \centering
        \includegraphics[trim=0cm 0cm 0cm -1cm, width=\linewidth]{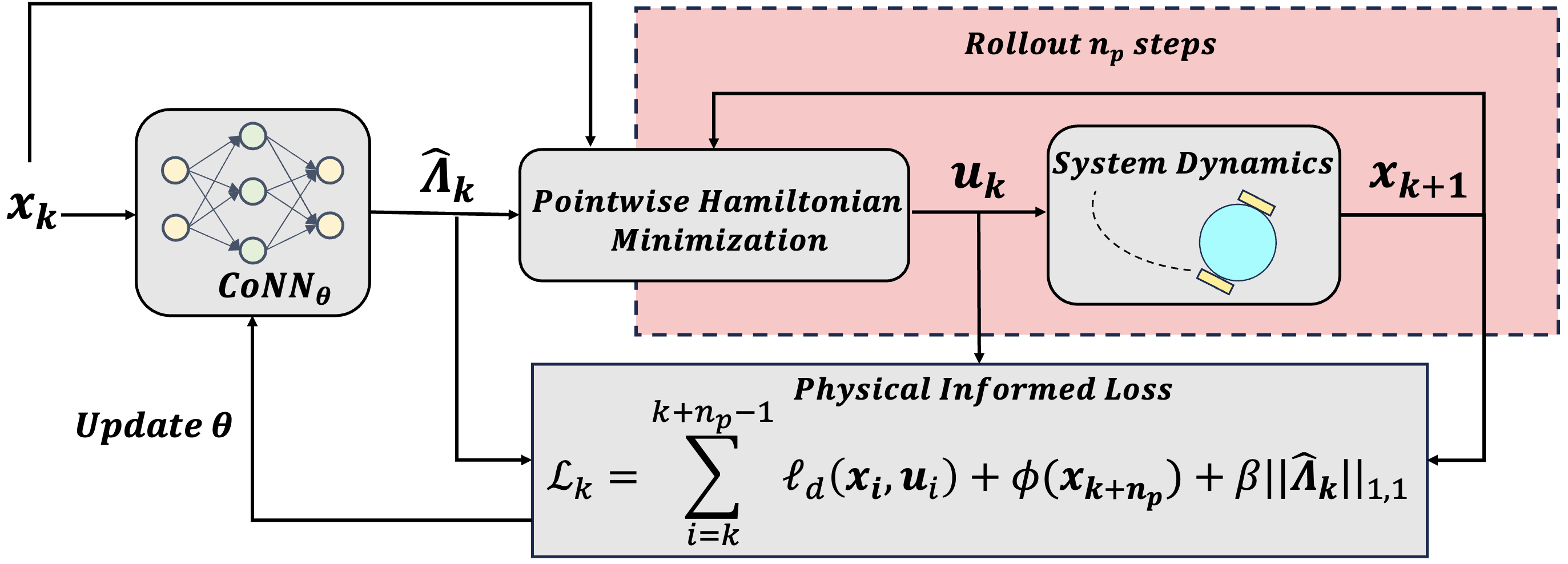}
        \caption[8pt]{Offline Training}
    \end{subfigure}
    \hfill
    \begin{subfigure}{0.44\textwidth}
        \centering
        \includegraphics[width=\linewidth]{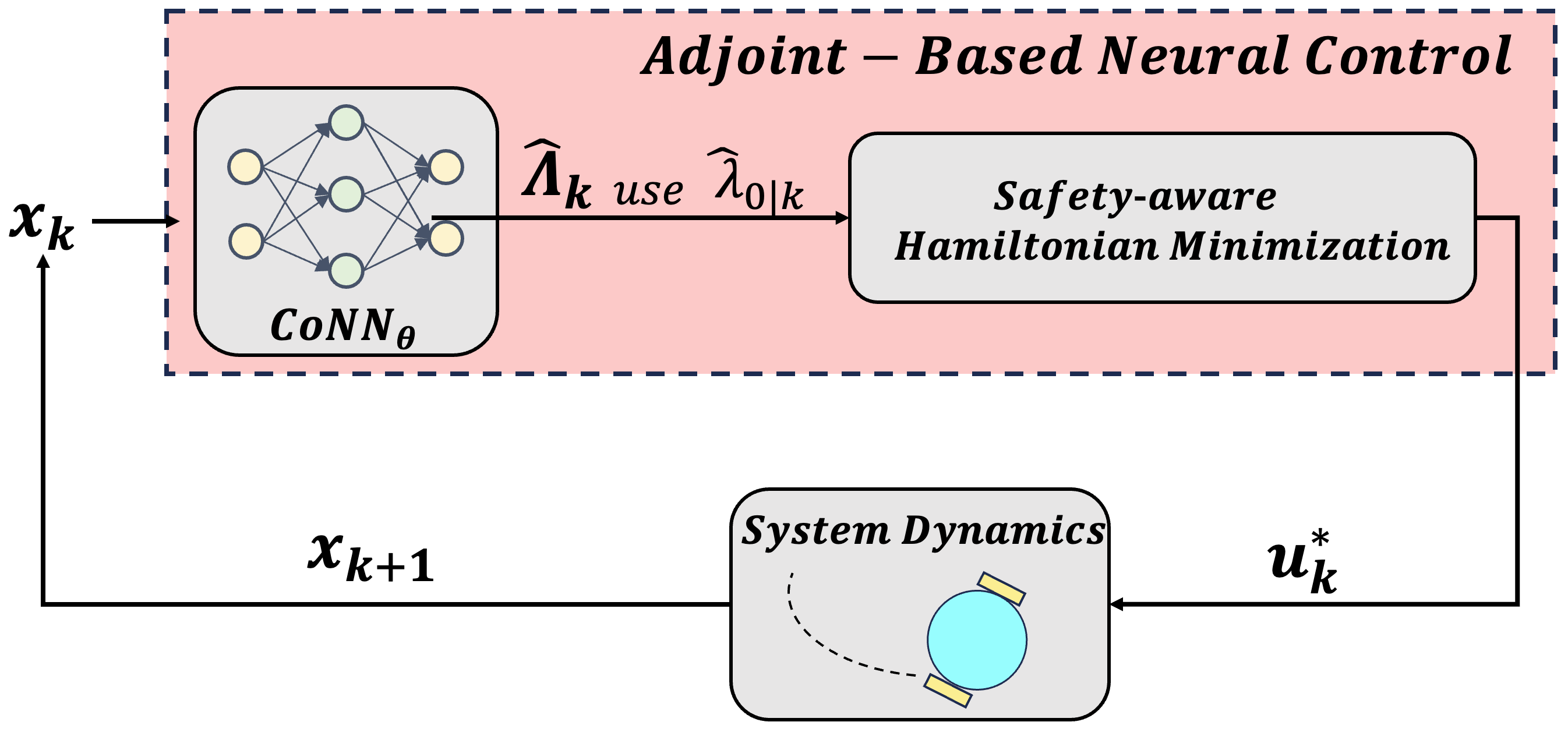}
        \caption[8pt]{Online Deployment}
    \end{subfigure}

    \caption{Adjoint-based Neural Regulator architecture. (a) During offline training, the CoNN predicts a finite-horizon co-state sequence from the current state. Pointwise Hamiltonian minimization and a horizon rollout are then used to define the physics-informed training loss. (b) During online deployment, the CoNN predicts a finite-horizon co-state sequence from the current state, and the first predicted co-state is used in a safety-aware constrained Hamiltonian minimization to compute the applied control.}
    \label{fig:architecture}
\end{figure*}

Motivated by the structure of PMP, we develop an adjoint-based neural regulator (ANR), illustrated in Fig.~\ref{fig:architecture}. The framework employs a co-state neural network (CoNN), following \cite{11312025}, to predict a finite-horizon co-state trajectory from the current state. The predicted co-state trajectory is then used to recover a nominal control sequence through pointwise Hamiltonian minimization. The resulting closed-loop rollout is used to define a training objective derived from the original optimal control cost. After training, the CoNN may be deployed in a feedback control scheme in which the predicted costate is now used to find the control input through a constrained Hamiltonian minimization in realtime. This yields a controller that preserves the fast online inference of learning-based methods while retaining the indirect optimality structure implied by PMP.

\subsection{Co-State Parameterization by CoNN}
As shown in Fig.~\ref{fig:architecture}, the CoNN maps an arbitrarily given state, denoted $\mathbf{x}_i$, to a co-state sequence over a prediction interval $[0,T]$: 
\begin{equation}
\hat{\boldsymbol{\Lambda}}_i
=
\mathrm{CoNN}_{\theta}(\mathbf{x}_i),
%\in\mathbb{R}^{n_p\times n},
\label{eq:conn_output}
\end{equation}
where $\theta$ denotes the trainable network parameters. The predicted sequence $\hat {\boldsymbol \Lambda}_i = \{\hat{\boldsymbol\lambda}_{0|i}, \hat{\boldsymbol\lambda}_{1|i}, \ldots, \hat{\boldsymbol\lambda}_{n_p-1|i}\}$, where

\[
\hat{\boldsymbol{\lambda}}_{k|i}
:=
\hat{\boldsymbol{\lambda}}( k\Delta t), ~\text{with } \mathbf{x}(0) = \mathbf{x}_i,
\quad k = 0,1,\ldots,n_p-1,
\] corresponds to co-state trajectory over the horizon $T=n_p\Delta t$, conditioned on the given state $\mathbf{x}_i$. In the implementation used here, the CoNN is realized as a fully connected feedforward neural network. We note that unlike in RL, the neural network is not used to imitate an action policy, but to parametrize the state-to-co-state mapping needed to construct a structured nominal controller.

\subsection{
Horizon Rollout for Training}%Nominal Control Recovery via Hamiltonian Minimization
Given the co-state sequence $\{\hat{\boldsymbol{\lambda}}_{k|i}\}_{k=0}^{n_p-1}$, we obtain the corresponding control input sequence by minimizing the Hamiltonian at each interval: 
\begin{align}
\hat {\mathbf{u}}_{k|i}^\star
&=
\arg\min_{\mathbf{u}\in\mathbb{R}^q}
H_d(\mathbf{x}_{k|i},\mathbf{u},\hat{\boldsymbol{\lambda}}_{k|i}),\label{eq:unconstrained_hamiltonian_min}
\end{align} 
For the quadratic cost in our objective (Eq. \eqref{eq:stage_terminal_cost}), this minimization may be explicitly computed, so that we may obtain the corresponding input and state sequences using 
\begin{align} 
    \hat {\mathbf{u}}_{k|i}^\star &= -\frac{1}{2} R^{-1}g^\top(\mathbf{x}_{k|i}) \hat{\boldsymbol{\lambda}}_{k|i}.\label{eq:unconstrained-optimal-u-expression}\\
    \mathbf{x}_{k+1|i} &= \int_{k\Delta t}^{(k+1)\Delta t}f(\mathbf{x}(t))
+
g(\mathbf{x}(t))\hat{\mathbf{u}}_{k|i}^\star~ dt
\label{eq:training_rollout_dyn} 
\end{align}
for $k=0,1,\dots,n_p-1$ with $\hat{\mathbf{x}}_{0|i} = \mathbf{x}_i$. This rollout process yields the trajectories $\{\hat{\mathbf{u}}_{k|i}\}_{k=0}^{n_p-1}$ and $\{\hat{\mathbf{x}}_{k|i}\}_{k=1}^{n_p}$
which are then used to construct the training loss.

\subsection{Training Objective}
The training objective consists of a three-term loss function: a stage-cost term, a terminal-cost term, and a co-state regularization term,
%The stage and terminal losses are given by 
\begin{equation}
\begin{gathered}[b]
\mathcal{L}_{\mathrm{stage}|i}
 =
\sum_{k=0}^{n_p-1}
\ell(\mathbf{x}_{k|i},\mathbf{u}_{k|i})=\sum_{k=0}^{n_p-1}
\mathbf{x}_{k|i}^\top Q \mathbf{x}_{k|i}
+
\mathbf{u}_{k|i}^\top R \mathbf{u}_{k|i},\\
\mathcal{L}_{\mathrm{terminal}|i}
=
\phi(\mathbf{x}_{n_p|i}) = \mathbf{x}_{n_p|i}^\top Q \mathbf{x}_{n_p|i}\\\mathcal{L}_{\mathrm{co\mbox{-}state}|i}
=
\beta \|\hat{\boldsymbol{\Lambda}}_i\|_{1,1},
\label{eq:loss_stage_general}
\end{gathered}
\end{equation}
where $\beta>0$ is a scalar hyperparameter and $\|\cdot\|_{1,1}$ denotes the entrywise $\ell_1$ norm. Together, these terms encourage the CoNN to produce the minimum-norm co-state predictions that induce cost-minimizing trajectories. The total training loss is given by
\begin{equation}
\mathcal{L}_{i}
=
\mathcal{L}_{\mathrm{stage}|i}
+
\mathcal{L}_{\mathrm{terminal}|i}
+
\mathcal{L}_{\mathrm{co\mbox{-}state}|i}.
\label{eq:total_loss}
\end{equation}
The CoNN parameters $\theta$ are optimized end-to-end by backpropagating through thehorizon rollout induced by Eqs.~\eqref{eq:unconstrained-optimal-u-expression}-\eqref{eq:training_rollout_dyn} for all $\mathbf{x}_{i}$ in a training domain $\mathcal{D}$.

\subsection{Safe Online Deployment via HOCBF-QP}
After offline training, the CoNN is deployed in feedback form using the current state $\mathbf{x}_k$, at each time $t_k$, to predict a finite-horizon co-state sequence \(
    \hat{\boldsymbol{\Lambda}}_k
=
\mathrm{CoNN}_{\theta}(\mathbf{x}_k)
=
\{\hat{\boldsymbol{\lambda}}_{0|k},\hat{\boldsymbol{\lambda}}_{1|k},\ldots,\hat{\boldsymbol{\lambda}}_{N-1|k}\},
\)
and the first predicted co-state $\hat{\boldsymbol{\lambda}}_{0|k}$ is used for control computation. The applied control is obtained by solving the following safety-aware constrained Hamiltonian minimization based on the HOCBF in Eq. \eqref{eq:hocbf_admissible_set}:
\begin{subequations}\label{eq:online_safe_control}
\begin{align}
\mathbf{u}_k^\star
&=
\arg\min_{\mathbf{u}}
\left(
\mathbf{u}^\top R \mathbf{u}
+
\hat{\boldsymbol{\lambda}}_{0|k}^{\top} g(\mathbf{x}_k)\mathbf{u}
\right), \\
\text{s.t.}\quad
& \mathbf{u}\in\mathcal{U}, \\
& \ddot h_i(\mathbf{x}_k,\mathbf{u})
+ k_{1,i}\dot h_i(\mathbf{x}_k)
+ k_{0,i} h_i(\mathbf{x}_k)
\ge 0, \label{eq:2ndorderHOCBF} \\
& \text{for}\qquad i=1,\dots,m. \notag
\end{align}
\end{subequations}
This yields a control law that directly incorporates the predicted co-state, actuator constraints, and safety constraints at each sampling instant.

\section{CASE STUDY AND RESULTS}\label{sec-example}

To evaluate the ANR's performance, we consider a unicycle robot with dynamics
\begin{equation}
\dot x = v\cos\theta,\qquad
\dot y = v\sin\theta,\qquad
\dot\theta = \omega,\qquad
\dot v = a,
\label{eq:unicycle}
\end{equation}
where \(\mathbf{x} = [x_p,\, y_p,\, \theta,\, v]^\top\) denotes the state and \(\mathbf{u} = [a,\, \omega]^\top\) the control input.
The objective is to steer the robot to a desired reference while satisfying actuator bounds and safety constraints. The control objective is of the form given in Eq.~\eqref{eq:background_ocp_formulation} with running cost \(
\ell(\mathbf{x},\mathbf{u})\) and terminal cost $\phi(\mathbf{x}(t_f))$. 

The admissible input set is chosen as \(
\mathcal U
=
\{(a,\omega): -1 \le a \le 1,\; -4 \le \omega \le 4\}\). The robot is constrained by arbitrarily placed circular obstacles in the $x-y$ plane.

Offline, the CoNN is trained end-to-end to map the current state (or tracking error), without any obstacles, to a finite-horizon co-state trajectory with length $n_p = 30$. During deployment, the first predicted co-state is used to determine the control action, while safety is enforced separately through a CBF filter. For each circular obstacle, we define
\begin{equation}
h_i(\mathbf{x}) = (x_p-x_{o,i})^2 + (y_p-y_{o,i})^2 - r_i^2,
\label{eq:hi}
\end{equation}
where \((x_{o,i},y_{o,i})\) is the obstacle center and \(r_i\)  is the effective safety radius. In this problem, the steering input $\omega$ does not appear in the first derivative of the barrier function, hence, a HOCBF, as given in \eqref{eq:2ndorderHOCBF}, is used to describe the admissible control set. Defining
\[
P_x=x_p-x_{o,i},\quad P_y=y_p-y_{o,i},\quad
s=P_x\cos\theta+P_y\sin\theta.
\]
we obtain,
\begin{align}
\dot h_i &= 2vs,\\
\ddot h_i &= 2v^2 + 2as + 2v\omega(-P_x\sin\theta + P_y\cos\theta).
\end{align}
Thus, the applied control is computed using the safety-aware constrained Hamiltonian minimization given in Eq.~\eqref{eq:online_safe_control}. Hence, the learned adjoint defines the control action, while the HOCBF-QP enforces input and safety constraints online.

\footnote{\thanks{$^{1}$We additionally evaluated SQP-RTI variant of NMPC. Although SQP-RTI is often preferred for real-time NMPC applications because of its lower computational cost, our SQP-RTI implementation exhibited reduced robustness. In particular, SQP-RTI showed noticeable control chattering and was more prone to early termination or failure to satisfy the terminal convergence criterion (see our \href{https://github.com/AI-Sys-Lab/Adjoint-based-Neural-Regulator.git}{github} page). Consequently, we use the \textit{ipopt} NMPC implementation for the benchmark results reported in this paper.}}

In the following subsections, we compare the performance of ANR with those of RL and NMPC as benchmarks under similar simulation scenarios\textsuperscript{1}. RL was carried out based on the SAC algorithm in \cite{9928337}. For NMPC, the \textit{ipopt} optimization solver is used for solving the nonlinear programming in \textit{CasADi}. %Note that while \textit{sqp} solvers are typically more efficient, our experiments showed them to be more prone to early termination.
Our results showed that while RL operates at the fastest in runtime, it generalizes poorly outside the training domain. By comparison, the ANR operates at slightly higher computational speeds but exhibits a generalizable performance that is similar to NMPC, including in cases with unseen initial states and references, as well as with or without obstacles. All simulations were performed on a desktop computer equipped with an Intel Core i9-14900K CPU and an NVIDIA GeForce RTX 4080 SUPER GPU.

\footnotetext{\thanks$^{2}$Absolute convergence error is the $\ell_1$-norm of the final state error.}

\subsection{\textbf{Obstacle-Free Scenarios}}
Here, we evaluate all three controllers (ANR, RL and NMPC) in three different obstacle-free cases.

\subsubsection{\textbf{Case A: In-distribution evaluation with seen $\mathbf{x}(0)$}}\label{no_obs:case_A}

First, we consider the obstacle-free scenario, where initial states were randomized within the training domain, and the performance of ANR and RL was evaluated over 100 trials. Specifically, the training was carried out, for both, on a grid over $[-2,2]$ for $(x,y,\theta,v)$ with 10 samples per state dimension. The ANR and RL achieved 100\%  and 30\% success rates, respectively, using an absolute convergence error (ACE) of $\leq 0.6$ as the success criterion\textsuperscript{2}. In addition, RL produced a lower average state and control mean square derivative (MSD) $(0.03$ and $0.05)$ compared to ANR $(0.17$ and $1.81)$. Ostensibly, the RL's low performance is due to its limited generalization to unseen initial states, even within the training domain. Although this first testing utilized initial states from the same range, because they were continuously randomly selected, many evaluation points were not encountered explicitly during training, hence, the poor performance on testing. 

For qualitative comparison, we selected an initial state where RL, ANR, and NMPC all show comparable ACE performance to show their respective navigation trajectories (see Case A in Fig.~\ref{fig:simulation1} and in Table ~\ref{tab:comparison}). RL operates at the fastest speed $(0.5 ms)$, compared to ANR $(1.2 ms)$ and NMPC $(347.7 ms)$. RL's slight computational advantage is expected because, during deployment, the SAC policy produces a nominal control input by direct policy evaluation, whereas ANR first predicts a costate trajectory and then solves a QP online based on the first predicted costate vector to recover the control action.

\subsubsection{\textbf{Case B: Out of Distribution Evaluation (Unseen $\mathbf{x}(0)$)}} \label{no_obs:case_B}
In this example, we evaluate generalization to initial conditions outside the training domain.  RL does not generalize reliably in this setting, whereas the proposed ANR shows good generalizability. The navigation trajectories of ANR, RL, and MPC are shown in Fig. \ref{fig:simulation1}. While MPC attains a lower ACE than ANR, it does so with a substantially higher computational cost, as summarized in Table~\ref{tab:comparison}.

\subsubsection{\textbf{Case C: Out of Distribution Evaluation with NonZero Reference $\mathbf{x}_{\mathrm{ref}}$}} \label{no_obs:Case_C}
\begin{figure}[h!]
    \centering
    \includegraphics[trim=0cm 0cm 0cm 0cm, width=\columnwidth]{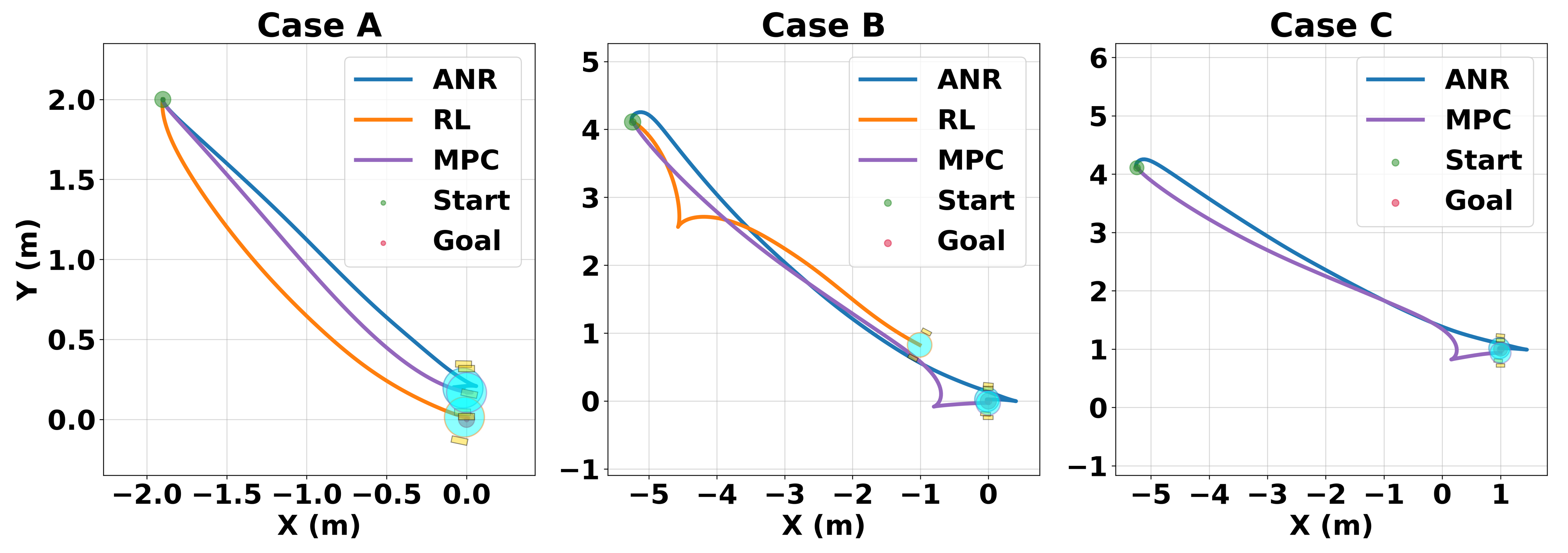}
    \caption[8pt]{\textbf{Without Obstacles}: Comparison of the representative initial state navigation trajectories.}
    \label{fig:simulation1}
\end{figure}

In this case, we used the same out-of-distribution initial state as in ~\ref{no_obs:case_B}, but changed the objective to track a nonzero reference, \(
\mathbf{x}_{\mathrm{ref}} = [1.00,\;1.00,\;0.00, \; 0.00]^\top.\)
Accordingly, the error state \((\mathbf{x}_k - \mathbf{x}_{\mathrm{ref}})\) was used as the input to the CoNN \cite{11312025}. We do not consider RL in this case study since it already showed poor performance outside the training domain in the earlier case.
Again, our ANR shows the capability of reaching the desired goal, as shown in Fig.~ \ref{fig:simulation1}, with comparable performance to MPC and at a much faster computational speed, as summarized in Table \ref{tab:comparison}.

\begin{table}[h]
    \centering
    \renewcommand{\arraystretch}{1.2} % Increase row height for better readability
    \begin{tabular}{c c c c}
        \toprule
        \textbf{\makecell{Performance \\ Metric}} & 
        \textbf{Case A} & 
        \textbf{Case B} & 
        \textbf{Case C} \\
        \midrule
        \textbf{\makecell{MSD \\ (State)}} & \makecell{ANR: 0.21 \\ \textit{RL  : 0.05}\\ MPC: 0.22 }& \makecell{ANR: 0.82 \\ \textit{MPC: 0.57}} & \makecell{ANR: 0.80 \\\textit{ MPC: 0.59}} \\
        \midrule
        \textbf{\makecell{MSD \\ (Control Input)}}& \makecell{ANR: 2.37 \\ \textit{RL  : 0.07}\\ MPC: 2.64}& \makecell{\textit{ANR: 3.02} \\ MPC: 5.77} & \makecell{\textit{ANR: 3.06} \\ MPC: 6.94} \\
        \midrule        
        \textbf{\makecell{Absolute \\ Convergence Error}} & \makecell{ANR: 0.26 \\ RL  : 0.25 \\ \textit{MPC: 0.17} }& \makecell{ANR: 0.14 \\ \textit{MPC: 0.04}} & \makecell{ANR: 0.13 \\ \textit{MPC: 0.07}} \\
        \midrule
        \textbf{\makecell{Computational \\ Speed $(ms)$}} & \makecell{ANR: 1.2 \\ \textit{RL   : 0.5}\\ MPC: 349.5}& \makecell{\textit{ANR: 1.2} \\ MPC: 370.1} & \makecell{\textit{ANR: 1.2} \\ MPC: 378.9} \\
        \bottomrule
    \end{tabular}
    \caption{ANR, RL, and MPC performance comparison without obstacles. \textit{Italicized} entries indicate better performance.}
    \label{tab:comparison}
    \vspace{-0.3cm} % Adjust space below the table if necessary
\end{table}

\subsection{\textbf{Obstacles-Laden Scenarios}}
Here, we evaluate all three controllers (ANR, RL, and NMPC) in cases with obstacles.
\subsubsection{\textbf{Case A: In distribution Evaluation with seen $\mathbf{x}(0)$}}

\begin{figure}[h!]
  \begin{subfigure}[b]{1\columnwidth}
   \begin{center}
      \includegraphics[trim=0cm 0cm 0cm 0cm, width=1\textwidth]{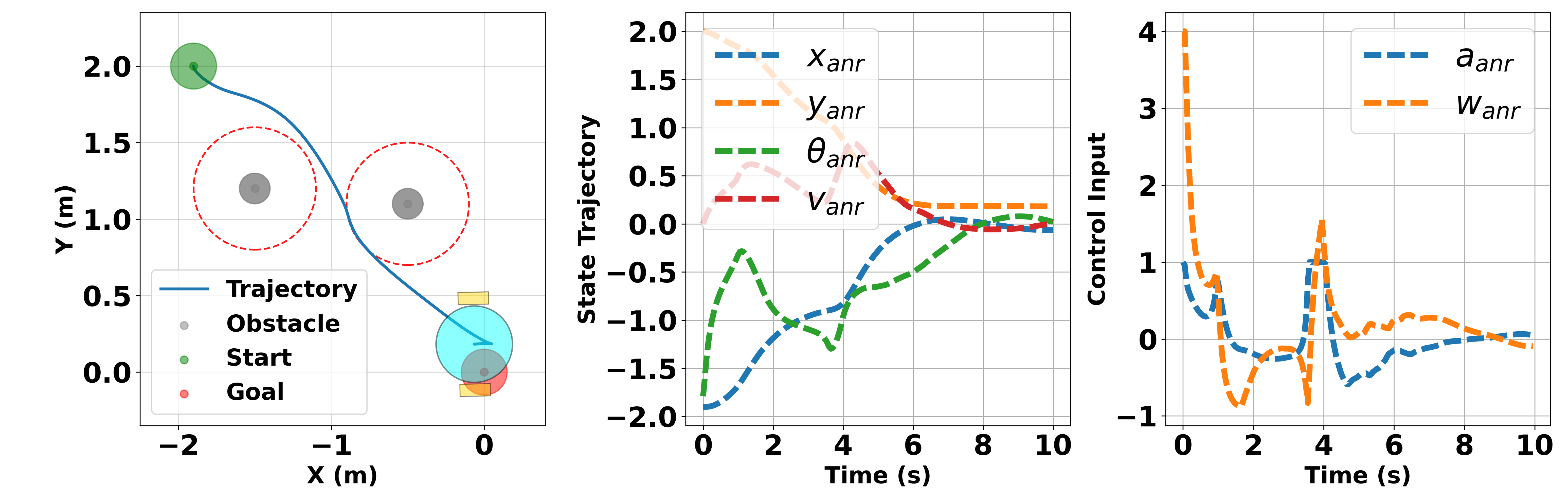}
      \caption[8pt]{ANR navigation, state, and control input trajectories.}
   \end{center}
  \end{subfigure}
  \begin{subfigure}[b]{1\columnwidth}
   \begin{center}
      \includegraphics[trim=0cm 0cm 0cm 0cm, width=1\textwidth]{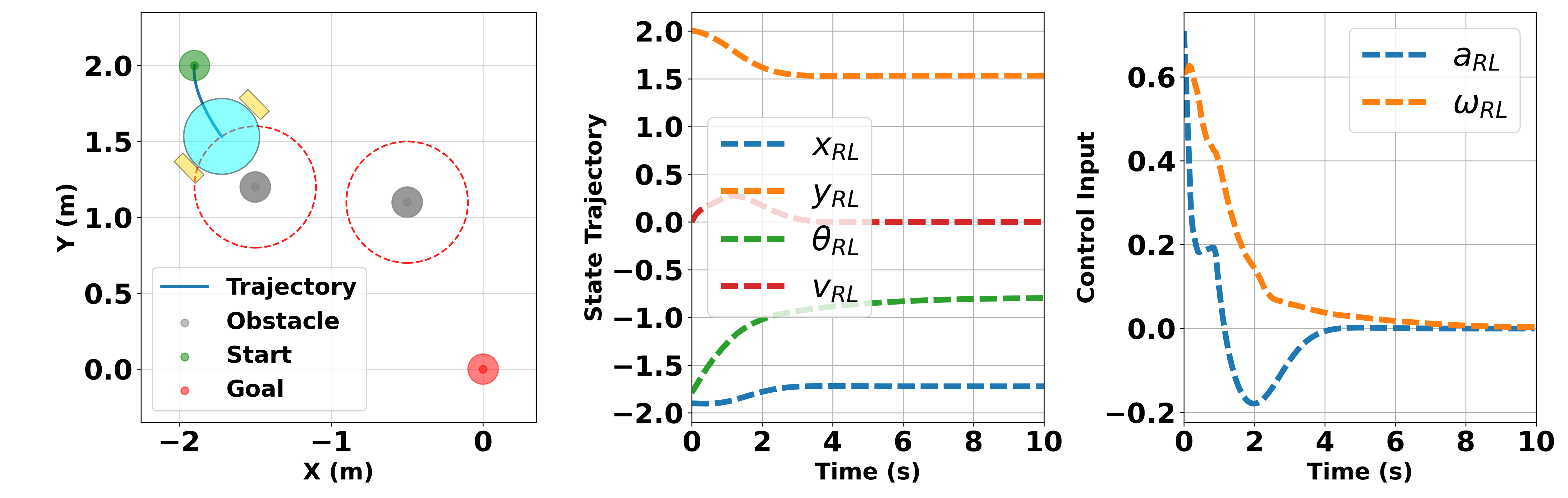}
      \caption[8pt]{RL navigation, state, and control input trajectories.}
   \end{center}
  \end{subfigure}
  \begin{subfigure}[b]{1\columnwidth}
   \begin{center}
      \includegraphics[trim=0cm 0cm 0cm 0cm, width=1\textwidth]{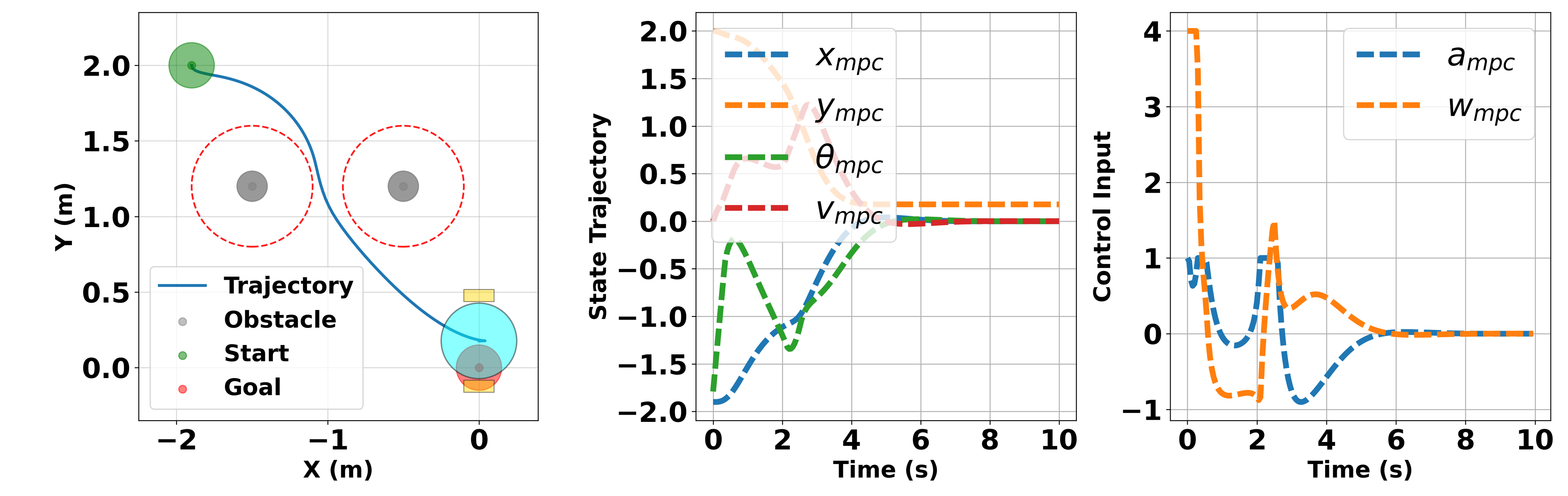}
      \caption[8pt]{MPC navigation, state, and control input trajectories.}
   \end{center}
  \end{subfigure}  
\caption[8pt]{In distribution comparison between ANR, RL, and MPC with obstacles}
\label{fig:simulation2}
\end{figure}

We first consider a constrained case with the same initial condition as the representative initial state used in ~\ref{no_obs:case_A}. We randomized the locations of two obstacles within the same domain. Using the same success criterion defined in ~\ref{no_obs:case_A}, performance was evaluated over 100 trials. ANR and RL achieved success rates of 82\% and 38\%, respectively. To illustrate the qualitative difference between the controllers, Fig.~\ref{fig:simulation2} shows a representative obstacle configuration, together with the corresponding state and control input trajectories, in which ANR reaches the target, whereas RL fails. We also included MPC for comparison.

\subsubsection{\textbf{Case B: Out of Distribution Evaluation (Unseen $\mathbf{x}(0)$})\label{with_obs:Case_B}}
\begin{figure}[h!]
  \begin{subfigure}[b]{1\columnwidth}
   \begin{center}
      \includegraphics[trim=0cm 0cm 0cm -1cm, width=0.96\textwidth]{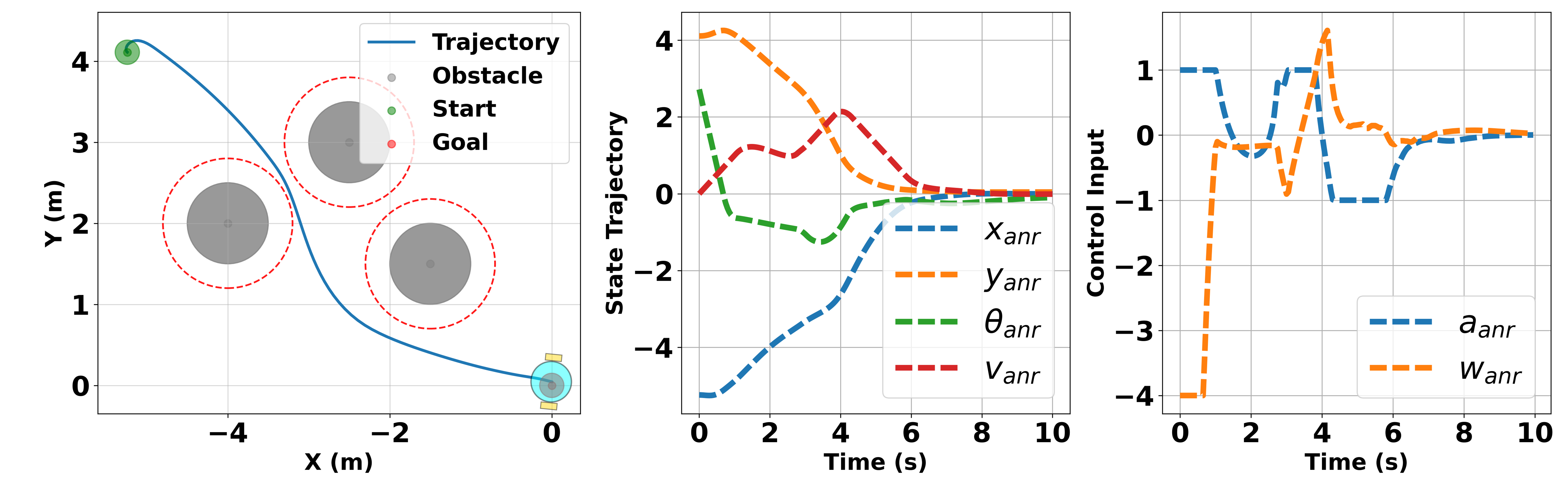}
      \caption[8pt]{ANR navigation, state, and control input trajectories.}
   \end{center}
  \end{subfigure}
  \begin{subfigure}[b]{1\columnwidth}
   \begin{center}
      \includegraphics[trim=0cm 0cm 0cm 0cm, width=0.96\textwidth]{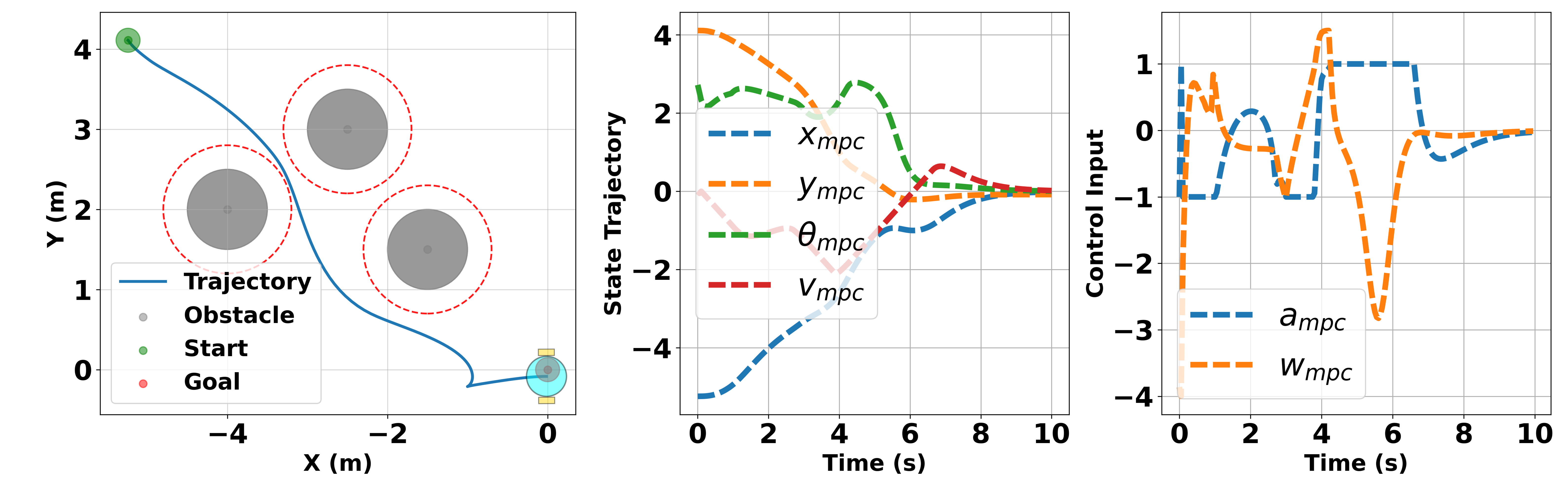}
      \caption[8pt]{MPC navigation, state, and control input trajectories.}
   \end{center}
  \end{subfigure}  
\caption[8pt]{Out of distribution comparison between ANR and MPC with obstacles.}
\label{fig:simulation3}
\end{figure}

For this case, we now consider constrained navigation with three densely-placed obstacles to evaluate how well our ANR generalizes outside the training domain with obstacles present.  Since RL already failed in the obstacle-free out-of-distribution setting, it is not included here; the results are summarized in Table \ref{tab:comparison2}. Fig.~\ref{fig:simulation3} shows the resulting navigation, state, and control input trajectories of both ANR and MPC.
To further evaluate ANR robustness, we selected an initial state outside the training domain and, in 100 trials, we randomized the obstacle locations within the feasible region between the start and goal while enforcing non-overlap and minimum separation. In this setting, ANR achieved a 98\% success rate.

\subsubsection{\textbf{Case C: Out of Distribution Evaluation with NonZero Reference $\mathbf{x}(0)$}}
\begin{figure}[h!]
  \begin{subfigure}[b]{1\columnwidth}
   \begin{center}
      \includegraphics[trim=0cm 0cm 0cm 1.5cm, width=1\textwidth]{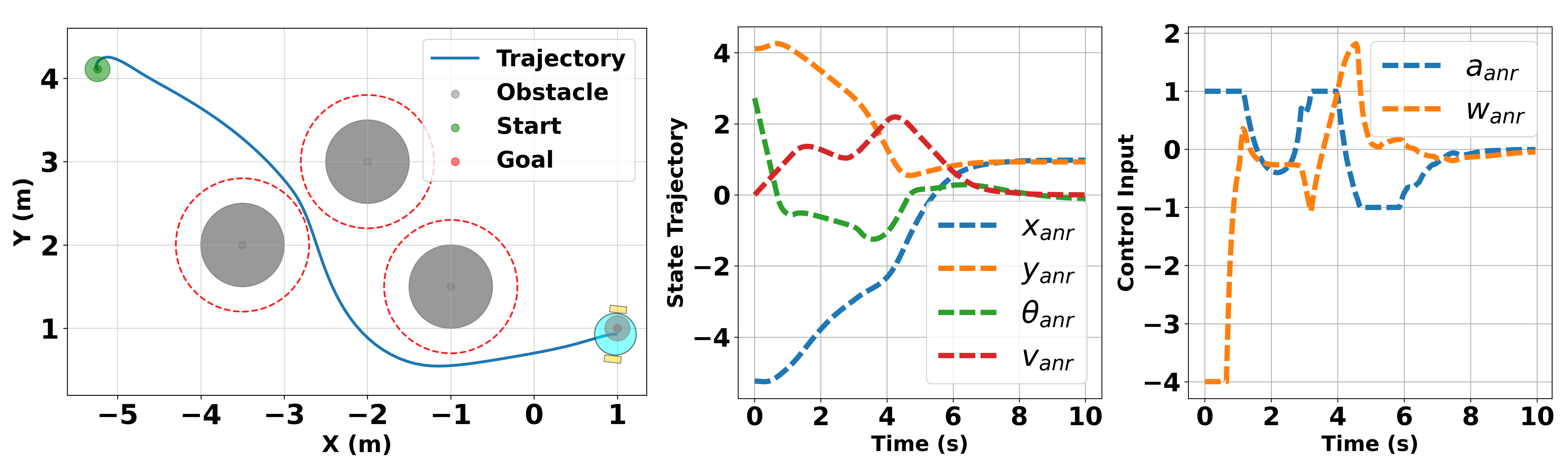}
      \caption[8pt]{ANR navigation, state, and control input trajectories.}
   \end{center}
  \end{subfigure}
  \begin{subfigure}[b]{1\columnwidth}
   \begin{center}
      \includegraphics[trim=0cm 0cm 0cm 0cm, width=1\textwidth]{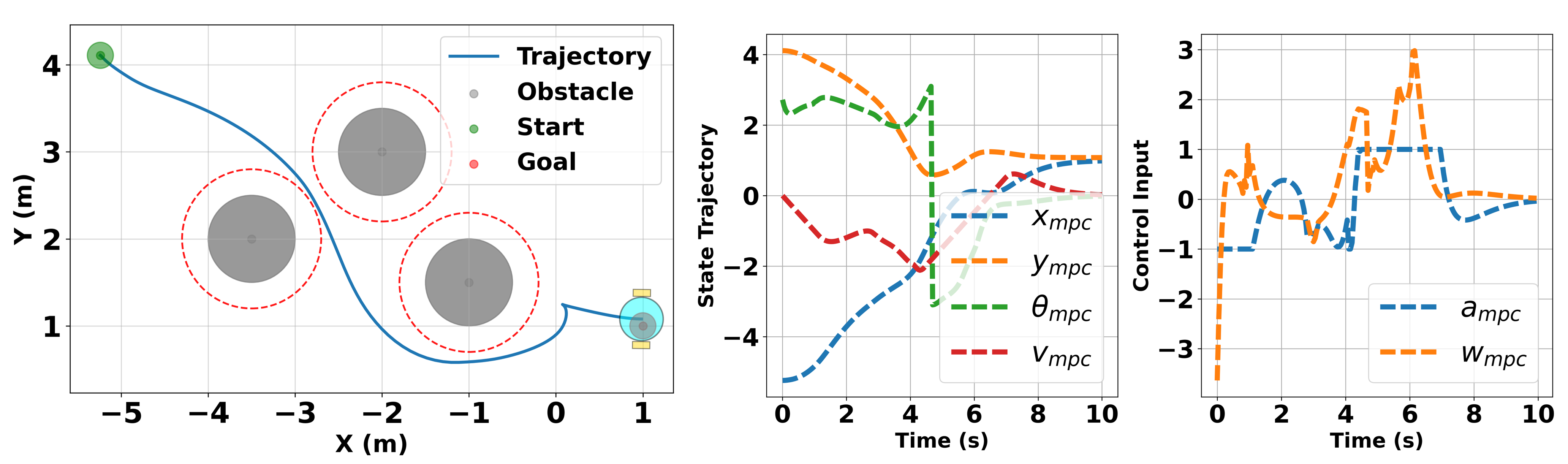}
      \caption[8pt]{MPC navigation, state, and control input trajectories.}
   \end{center}
  \end{subfigure}  
\caption[8pt]{Comparison between ANR and MPC with obstacles for out-of-distribution with nonzero reference.}
\label{fig:simulation4}
\end{figure}

As explained in ~\ref{no_obs:Case_C}, the objective to track a nonzero reference, and we do not consider RL in this case study since it already showed poor performance outside the training domain. Using the same out-of-distribution initial state as in \ref{with_obs:Case_B}, we then consider a constrained navigation with three densely-placed obstacles to evaluate how well our ANR generalizes outside the training domain with non-zero reference. Again, our ANR shows comparable results to MPC, at a much faster computational speed, as summarized in Table \ref{tab:comparison2}. Fig.~\ref{fig:simulation4} shows the navigation, state, and control input trajectories, with obstacles for both ANR and MPC.

\begin{table}[h]
\vspace{6pt}
    \centering
    \renewcommand{\arraystretch}{1.2} % Increase row height for better readability
    \begin{tabular}{c c c c}
        \toprule
        \textbf{\makecell{Performance \\ Metric}} & 
        \textbf{Case A} & 
        \textbf{Case B} & 
        \textbf{Case C} \\
        \midrule
        \textbf{\makecell{MSD \\ (State)}} & \makecell{\textit{ANR: 0.19} \\ MPC: 0.30 }& \makecell{ANR: 0.70 \\\textit{ MPC: 0.56}} & \makecell{\textit{ANR: 0.76} \\ MPC: 10.0} \\
        \midrule
        \textbf{\makecell{MSD \\ (Control Input)}}& \makecell{\textit{ANR: 3.79} \\ MPC: 5.09}& \makecell{\textit{ANR: 3.49} \\ MPC: 7.76} & \makecell{\textit{ANR: 3.97} \\ MPC: 9.05} \\
        \midrule
        \textbf{\makecell{Absolute \\ Convergence Error}} & \makecell{ANR: 0.28 \\ \textit{MPC: 0.18}}& \makecell{ANR: 0.16 \\ \textit{MPC: 0.09}} & \makecell{ANR: 0.20 \\\textit{ MPC: 0.13}} \\
        \midrule
        \textbf{\makecell{Computational \\ Speed $(ms)$}} & \makecell{\textit{ANR: 1.4} \\ MPC: 803.8 }& \makecell{\textit{ANR: 1.4} \\ MPC: 1206.7} & \makecell{\textit{ANR: 1.4} \\ MPC: 1228.9} \\
        \bottomrule
    \end{tabular}
    \caption{ANR and MPC performance comparison with obstacles. \textit{Italicized} entries indicate better performance.}
    \label{tab:comparison2}
    \vspace{-0.3cm} % Adjust space below the table if necessary
\end{table}

\section{Conclusion}\label{sec-conclusion}

In this paper, we utilized an ANR framework based a CoNN that incorporates HOCBF constraints into the online control optimization, such that actuator limits and safety constraints are both enforced. The resulting ANR-HOCBF controller retains the low online computational cost of the CoNN framework while extending its applicability to safety-critical control problems. We applied the proposed approach to a unicycle robot navigating in an unknown environment under real-time safety constraints. Results show that the proposed approach achieves performance comparable to MPC at a much lower computational cost while generalizing more effectively than RL to unseen initial states and references. Future work will consider dynamic obstacles and higher-dimensional systems.

\bibliographystyle{IEEEtran}
\bibliography{ref}

\end{document}